\documentclass[aps, prl, reprint, groupedaddress, footinbib, showpacs]{revtex4-1}

\usepackage{amsmath}
\usepackage{amssymb}
\usepackage{mathtools}
\usepackage{xcolor}
\usepackage{graphicx}
\usepackage{xfrac}
\usepackage{xcolor}
\usepackage{subfigure}
\usepackage{bm}
\usepackage{hyperref,hypcap}
\hypersetup{
	colorlinks,
	citecolor=red,
	linkcolor=blue,
	urlcolor=blue}

\begin{document}
%Title of paper
\title{Anomalous Drag in Double Bilayer Graphene Quantum-Hall Superfluids}
\author{Ming Xie}
\affiliation{Department of Physics, The University of Texas at Austin, Austin, TX 78712, USA}
\author{A. H. MacDonald}
\affiliation{Department of Physics, The University of Texas at Austin, Austin, TX 78712, USA}

\date{\today}

\begin{abstract}
Semiconductor double-layers in the quantum Hall regime tend to have
superfluid exciton condensate ground states when the total filling factor is an odd integer,
provided that the Landau orbitals at the Fermi level in the two layers have the same orbital character.
Since the $N=0$ Landau level of bilayer graphene contains 
states with both $n=0$ and $n=1$ orbital character, the physics of double bilayers falls outside 
previously studied cases.  We show that the superfluid phase stiffness 
vanishes in double bilayer graphene when $n=0$ and $n=1$ orbitals states are 
degenerate in one of the layers, even though the gap for charged excitations remain large, 
and speculate that this property is behind the recent discovery
of strong anomalous drag near a $n=0/1$ degeneracy point.  
\end{abstract}

%\pacs{to be filled} 
%Use showkeys class option if keyword display desired
\maketitle

\emph{Introduction.}---\nobreakdash
Because of kinetic energy quantization, states with broken symmetries, often referred to 
generically as quantum Hall ferromagnets, are common
whenever several Landau levels are close to degeneracy and the total Landau level filling 
factor is an integer.  In semiconductor double-layers, exciton condensate states characterized
by spontaneous interlayer phase coherence, counterflow superfluidity 
\cite{Fertig1989, Wen1992, MacDonald1990a, MacDonald1990b, Eisenstein2004, Tutuc2004}, 
and other unique anomalous transport properties, appear when 
the total Landau level filling factor is an odd integer and the
states at the Fermi level have the same orbital character in both layers \cite{Jungwirth2000}.
In this Letter we address the interesting case in which one or both sides of the 
double layer are formed by bilayer graphene (BLG) systems with a low energy 
$N=0$ Landau level.  The 8-fold degenerate $N=0$ Landau level multiplet of BLG is  
the direct product of real spin, valley pseudospin, and $n=0,1$ orbital pseudospin doublets.
We show that in this case the exciton superfluid phase stiffness vanishes
while the charged excitation gap remains finite 
whenever the orbital doublet in one or 
both layers approaches exact degeneracy.  Because a large negative drag 
is expected whenever the resistivity for collective counterflow transport
exceeds the resistivity for parallel quasiparticle charge flow, we argue that this property 
is behind recent observation of large negative drag
when the $n=0$ orbital energy in one bilayer approaches the $n=1$ orbital energy from above.
When the $n=1$ orbital energy is below the $n=0$ orbital energy 
states with long period spatial modulation of orbital character can occur.  

\begin{figure}[h!]
	\centering
	\includegraphics[width=0.999\columnwidth]{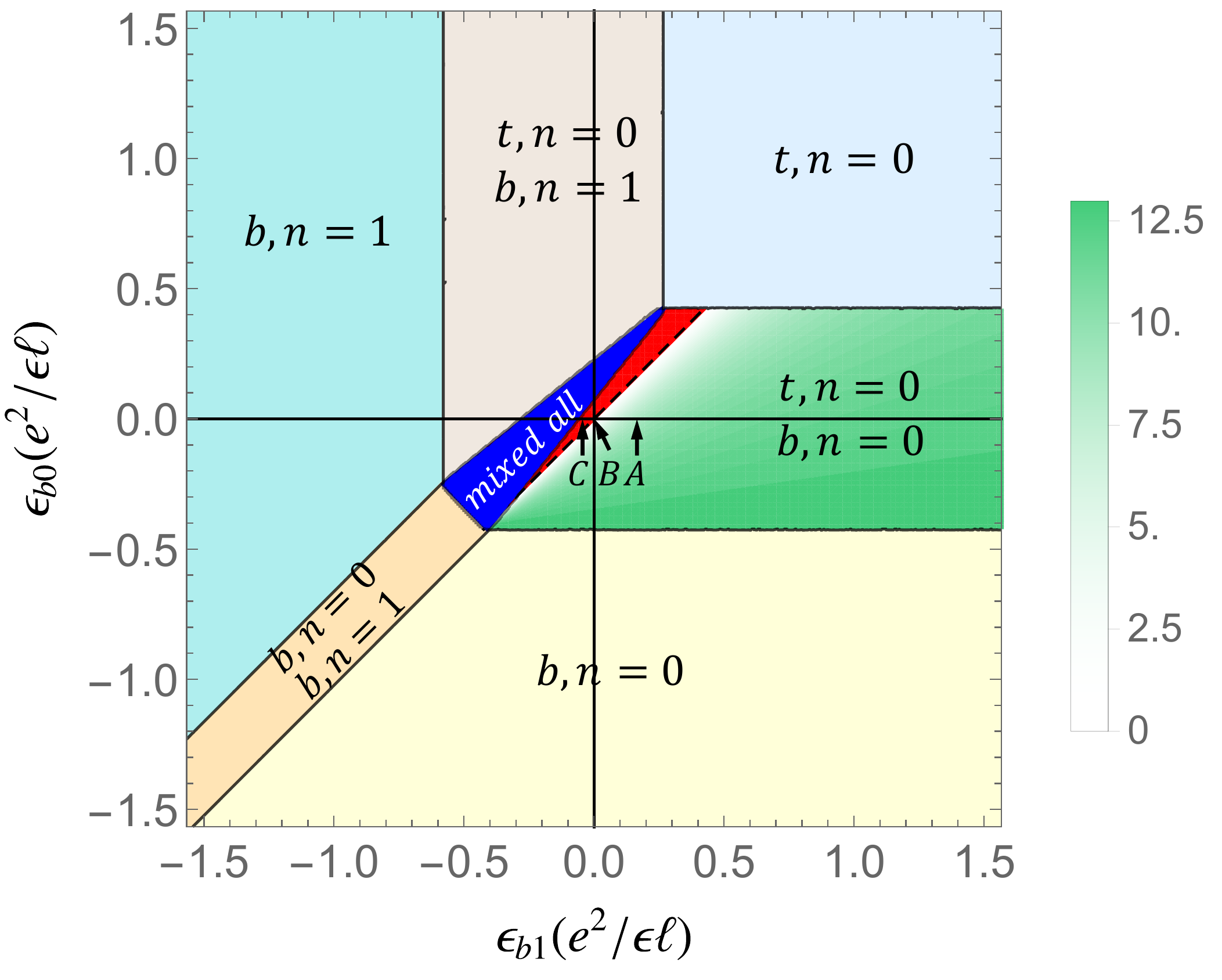}
	\caption{\label{fig:phase_diagram} 
		Phase diagram of bilayer graphene double layers
		as a function of the single particle energies, $\epsilon_{b0}$ and $\epsilon_{b1}$, of the 
		$n=0$ and $n=1$ orbitals in the bottom bilayer relative the energy of the $n=0$ orbital in the top layer.
		A bilayer exciton condensate (green region) forms when $\epsilon_{b0}$ is close to zero and 
		$\epsilon_{b1} > \epsilon_{b0}$. 
		We show that the superfluid phase stiffness of this condensate (quantified
		in $10^{-3} e^2/\epsilon\ell^3$ units by the brightness of the green region
		using the color scale on the right) 
		vanishes as $\epsilon_{b1}$ approaches $\epsilon_{b0}$ from above, and 
		predict that anomalous drag properties occur for very small superfluid densities.
		When $\epsilon_{b1} < \epsilon_{b0} \sim 0$ (red region)
		, we find that the ground state is characterized by  
		long period spatial modulation of the orbital character. 
		The orange and gray regions of the phase diagram 
		 exhibits Ising quantum Hall ferromagnetism.   In other regions of the phase diagram, the 
		electrons are completely layer polarized.}
\end{figure}

\emph{Bilayer Ground States.}---\nobreakdash
The phase diagram of a bilayer system with a $n=0$ orbital Landau level in the top 
layer and $n=0,1$ close to degeneracy in the bottom layer is illustrated in Fig.~\ref{fig:phase_diagram}.
To construct this phase diagram we have assumed that
both real spin and valley pseudospin are polarized, allowing us to focus on orbital pseudospin, 
and compared the energies of 
translationally invariant single Slater determinant full Landau level
states with arbitrary orbital content,
\begin{equation}
	|\Psi[z]\rangle =\prod_{X} \big(\sum_{\alpha,n}z_{\alpha n} \, \hat{c}^\dagger_{\alpha n, X} \big)|vac\rangle.
\label{Eq:ground_state}
\end{equation}
In Eq.~\ref{Eq:ground_state} $|vac\rangle$ is the vacuum state in which the Landau levels of interest 
are empty, $X$ is the guiding center label of the individual quantum states within a 
Landau level, $\hat{c}^\dagger_{\alpha n,X}$ is the creation operator for a state 
with layer index $\alpha=t, b$, and orbital index $n=0,1$, and the  
$z_{\alpha n}$ are the components of the ground state layer/orbital spinor
which satisfies the normalization constraint:
$\sum_{\alpha,n} \, |z_{\alpha n}|^2=1$.
Extrema of $E[z] = \langle \Psi[z] | {\cal H} | \Psi[z] \rangle$ are 
eigenstates of the mean field Hamiltonian,
\begin{eqnarray}
\mathcal{H}^{\rm MF}_{\alpha n, \beta n'} &=& 
% single particle + Hartree
\Big[ \epsilon_{\alpha,n} + \Big(\sum_{n_1}|z_{\alpha n_1}|^2-\frac{1}{2}\Big)\frac{d}{\ell}\Big] \delta_{\alpha\beta}\delta_{nn'} \nonumber \\
% Exchange
&+& \sum_{n_1,n_2} X^{\alpha\beta}_{nn_1n_2n'} z_{\alpha n_1}z^*_{\beta n_2}.
\label{Eq:Hmeanfield}
\end{eqnarray}
When projected to the valence Landau levels of interest the many-body Hamiltonian $\cal{H}$ includes 
only $X$-independent single-particle energies and Coulomb interactions.
In Eq.~\ref{Eq:Hmeanfield} $\epsilon_{\alpha,n}$ is an orbital-dependent single particle energy, which 
can be tuned by adjusting gate voltages \cite{McCann2006}
and includes a self-energy contribution\cite{Shizuya2010} 
due to exchange interactions with occupied Dirac sea Landau levels.
The second term in square brackets in Eq.~\ref{Eq:Hmeanfield}  is the Hartree self-energy, the final
term is the exchange self-energy,
$d$ is spacing between the two  bilayers,
and energies are in units of $e^2/(\epsilon \ell)$,
where $\epsilon$ is the background dielectric constant.
The exchange integrals in Eq.~\ref{Eq:Hmeanfield} 
are given by \cite{Barlas2010, Cote2010}
\begin{align}
	X^{\alpha\beta}_{nn_1n_2n'} = 
	\int \frac{d^2\bm{q}}{(2\pi)^2} V_{\alpha\beta}(\bm{q}) 
	F_{nn_1}(\bm{q}) F_{n_2n'}(-\bm{q})
\end{align}
where $V_{\alpha\beta}(\bm{q})= 2 \pi e^2/(\epsilon |\bm{q}|) \exp(-qd_{\alpha\beta})$, 
$F_{n'n}(\bm{q})$ is an orbital dependent form factor, and 
$d_{\alpha\beta}=d(1-\delta_{\alpha\beta})$.
For many values of 
$\epsilon_{b1}-\epsilon_{t0}$ and $\epsilon_{b0}-\epsilon_{t0}$, there is more than one self-consistent 
solutions of these mean-field equations, and the ground state must be determined by comparing 
extremal energies.

\emph{Phase stiffness of the Monolayer/Bilayer Exciton Condensate.}---\nobreakdash
Corrections can be added to these mean-field states by deriving a 
quantum fluctuation Hamiltonian.
For this purpose we expand the layer/orbital spinors in terms of 
the mean-field theory eigenspinors:
\begin{align}
	|m,X\rangle = \sum_{\alpha, n} a^m_{\alpha n} |\alpha, n, X\rangle,\ \ m=0,1,2
\end{align}
where $a^m_{\alpha n}$ is the eigenvector of the self-consistent mean-field Hamiltonian,
the label $m=0$ is reserved for the occupied lowest energy spinor,
and $m=1,2$ for the two unoccupied excited state spinors. 
Note $a^m_{t1}$ is always zero because we assume that the $m=1$ orbital 
energy is close to degeneracy only in the bottom layer. 

The fluctuation energy functional is constructed by considering instantaneous Slater determinants
\begin{equation}
|\Psi[z] \rangle =\prod_{X} \big(\sum_{m=0}^2 z_{m, X}\hat{c}^\dagger_{m, X} \big)|vac\rangle,
\end{equation} 
with guiding-center dependent orbitals that contain
small admixtures of higher energy mean-field eigenspinors.
To second order in the fluctuation amplitudes 
$z_{1,X}$ and $z_{2,X}$, the fluctuation energy is specified by the kernel
\begin{equation}
	\mathcal{K}_{ij}(X-X') = \frac{\partial^2 \mathcal{E}[Z]}{\partial Z_{i,X}\partial Z_{j,X'}}.
\label{Eq:fluctuationkernel}
\end{equation}
In Eq.~\ref{Eq:fluctuationkernel} we have defined $Z_{i,X}=\{z_{1,X}, z^*_{1,X}, z_{2,X}, z^*_{2,X}\}$.
Explicit forms for the total energy and for the fluctuation kernel are listed in the 
supplemental material\ref{supple}.
Because the elements of the fluctuation kernel depend only on the difference between guiding centers
they can be Fourier transformed and are most conveniently expressed in terms of 
$\mathcal{K}_{ij}(q)=\sum_X e^{iqX}\mathcal{K}_{ij}(X)$.

\begin{figure}[t]
	\centering
	\includegraphics[width=0.99\columnwidth]{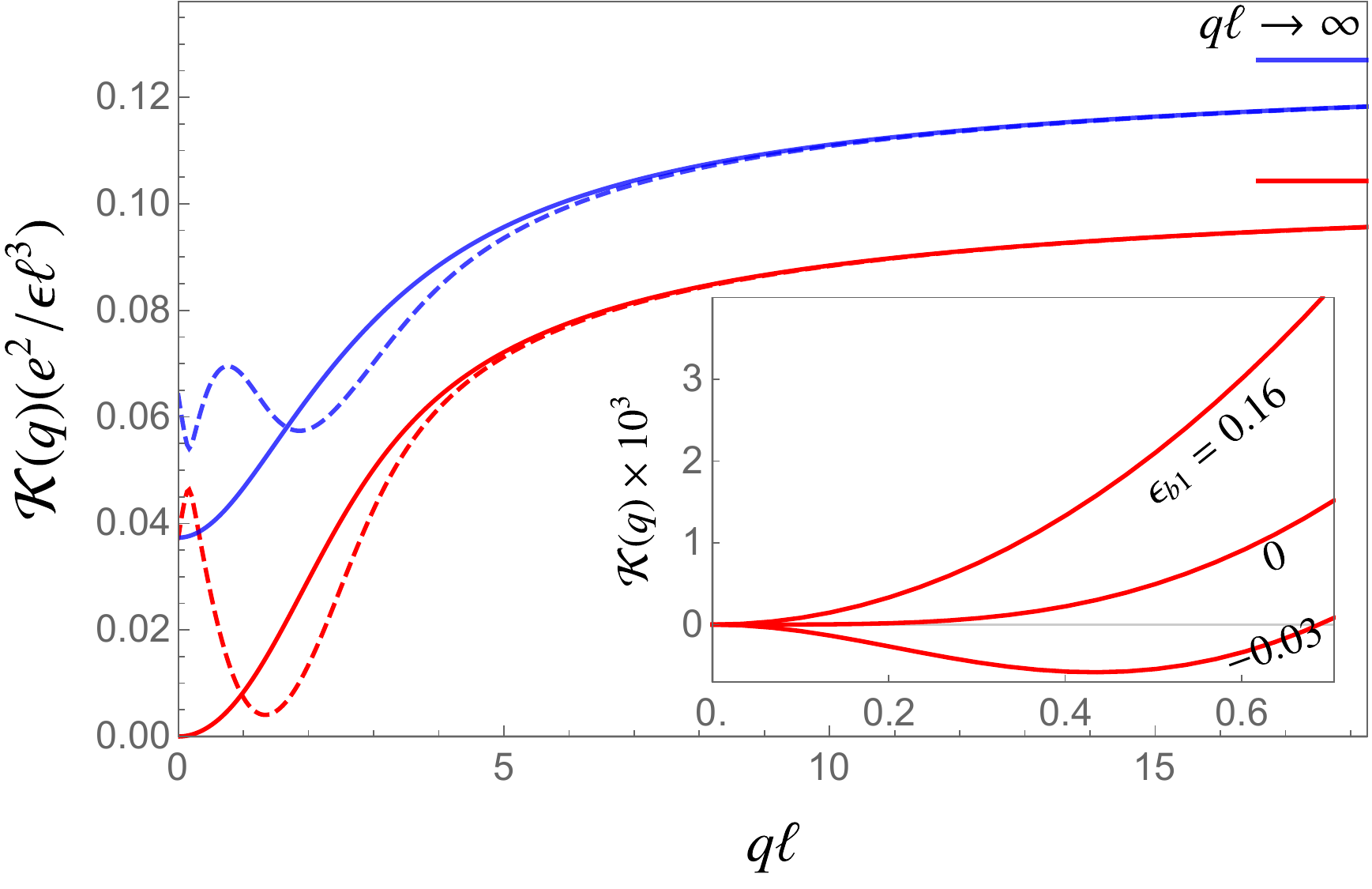}
	\caption{\label{fig:kernel} Phase (solid) and density (dashed) fluctuation eigenenergies of the
		stability kernel $\mathcal{K}(q)$. Here we set $\epsilon_{b0}=0$. 
		The main panel plots results for point A in the phase diagram Fig.\ref{fig:phase_diagram}.
		The inset shows a zoom in of the lowest energy branch for$q \to 0$ for points 
		A, B, and C, in Fig.\ref{fig:phase_diagram} at which $\epsilon_{b1}=0.016, 0, -0.03$ respectively.} 
\end{figure}

Our main interest here is in the green region of Fig.~\ref{fig:phase_diagram}, 
where the mean field ground state is an exciton condensate with 
spontaneous phase coherence between $n=0$ orbitals in the top and bottom layers,
empty $n=1$ orbitals, and counterflow superfluidity which we discuss at greater depth below.  
Concentrating first on the $\epsilon_{b0}=0$ line, along which the top and bottom layers have equal 
weight, the bilayer exciton condensate state remains stable at the mean-field level even 
when $\epsilon_{b1} < \epsilon_{b0}$ because the exchange energy between $n=0$ orbitals is 
larger than the exchange energies that involve $n=1$ orbitals.
For $\epsilon_{b1} - \epsilon_{b0} < - 0.07e^2/\epsilon\ell$, the dark blue 
region marked as ``\textit{mixed all}" in Fig. \ref{fig:phase_diagram}, 
the mean field theory spinors have non-zero projections onto all three orbitals.

In Fig.~\ref{fig:kernel} we plot  eigenvalues of the fluctuation kernel $\mathcal{K}_{ij}(q)$ 
of the $n=0$ bilayer exciton condensate state for $\epsilon_{b0}=0$ and  
several values of $\epsilon_{b1}$.
(The relationship of these eigenvalues to
the collective mode energies is explained in the supplemental material.)
At quadratic order, fluctuations that influence the interlayer 
phase (solid lines) and density balance (dashed lines) decouple.
If we choose the mean-field state to have the same phase in both layers, the 
basis functions for phase and density fluctuations are 
\begin{align}
	z_{1,X}^\pm = (z_{1,X} \pm z^*_{1,X})/\sqrt{2}, \notag\\
	z_{2,X}^\pm = (z_{2,X} \pm z^*_{2,X})/\sqrt{2}.
\end{align}
where the ``$+$'' sign corresponds to density and the ``$-$'' sign to phase modes.
The $i=1$ excited state is an antisymmetric $n=0$ bilayer state, 
and the $i=2$ excited state is a $n=1$ orbital state localized in the bottom layer.
The vanishing energy of the phase mode as $q \to 0$, reflects the continuously broken interlayer 
phase $U(1)$ symmetry.  The coefficient of $q^2$ in the phase mode energy is the 
superfluid phase stiffness, which is the key microscopic property of the
exciton condensate \cite{Yang1996} because it controls the Kosterlitz-Thouless phase transition 
temperature and the relationship between counterflow superfluid currents and 
interlayer phase gradients.  
Our calculations show that the superfluid phase stiffness decreases as 
$\epsilon_{b1} \to \epsilon_{b0}$ from above, vanishing along the 
$\epsilon_{b1} = \epsilon_{b0}$ black dashed line in Fig.~\ref{fig:phase_diagram}.  

Because of kinetic-energy quenching in a strong magnetic field, 
the superfluid phase stiffness
 of quantum Hall bilayer excitonic condensates is entirely due to electron-electron 
interactions, with no contribution from electron-hole pair kinetic energy.
To explain why the superfluid phase stiffness vanishes when 
$\epsilon_{b1} \to \epsilon_{b0}$, we perform an analytic Taylor 
series expansion of the phase subspace of the fluctuation kernel in powers of $q=|\bm{q}|$ for 
$\epsilon_{b0}=\epsilon_{t0}$ which yields
\begin{align}
\mathcal{K}^-(q) =
	\begin{pmatrix}
	f(d) q^2\ell^2 & -if(d)q\ell \\
	if(d)q\ell & (\epsilon_{b1}-\epsilon_{b0})+f(d) +\Delta(d)q^2\ell^2
	\end{pmatrix}.
\label{Eq:longwavelengthfluctuation}
\end{align}
Here 
$f(d) = -\tfrac{d}{4\ell}+\tfrac{1}{4}\sqrt{\tfrac{\pi}{2}}\big(1+\tfrac{d^2}{\ell^2}\big)e^{\tfrac{d^2}{2\ell^2}}\text{Erfc}\big(\tfrac{d}{\sqrt{2}\ell}\big)$, and $\Delta(d)$ are 
layer-separation dependent positive constants.
To second order in $q$, the lower eigenenergy of $\mathcal{K}^-(q) $ is 
\begin{align}
	E^-(q) = \frac{f(d) \, (\epsilon_{b1}-\epsilon_{b0})}{f(d) +  (\epsilon_{b1}-\epsilon_{b0})} \; q^2\ell^2,
\end{align}
which vanishes as $(\epsilon_{b1} \to \epsilon_{b0})$ from above
as illustrated in Fig.~\ref{fig:kernel}.
For $ \epsilon_{b1} - \epsilon_{b0} \gg f(d)$ the coefficient of $q^2\ell^2$ in Eq.~\ref{Eq:longwavelengthfluctuation},
which is the superfluid phase stiffness, approaches $f(d)$, its standard bilayer limit.

When the interlayer phase has a spatial gradient \cite{Radzihovsky2001}
nearby guiding centers are in different bilayer states, reducing the magnitude of 
the attractive exchange interaction.  Because the transverse orbital of the $n=1$ guiding 
center state is the derivative of the $n=0$ orbital, mixing $n=1$ orbitals relaxes the constraint 
that locks the Landau gauge guiding center label to the wavefunction maximum.  The 
structure of Eq.~\ref{Eq:longwavelengthfluctuation}, in which the same quantity $f(d)$ appears 
in both the  $(1,1)$ and $(2,2)$ diagonal and the $(1,2)$ and $(2,1)$ off-diagonal matrix elements,
reflects the property that the exchange energy cost of small phase gradients can be 
completely eliminated.  The superfluid phase stiffness is therefore finite only if there is a 
single-particle energy cost of mixing $n=1$ orbitals into the ground state.

\emph{Negative Drag.}---\nobreakdash
Even though the superfluid phase stiffness approaches zero as $\epsilon_{b1} \to \epsilon_{b0}$,
the condensation energy of the excitonic state and its charge excitation gap remain large.
In this case we expect that the excitonic character of the many-particle ground state will remain intact 
in this region of the phase diagram.  The small superfluid phase stiffness
then implies a broad region of 
temperature in which charged excitations are dilute and the
Kosterlitz-Thouless transition temperature of the exciton fluid, which is bounded by 
the zero temperature superfluid density $\rho_0$, is substantially exceeded:
\begin{equation} 
k_{B} T_{KT}  < \frac{\pi}{2} \rho_0 = \frac{\pi}{2} \;  \frac{f(d) \, (\epsilon_{b1}-\epsilon_{b0})}{f(d) +  (\epsilon_{b1}-\epsilon_{b0})}.
\end{equation}
($f(d)\approx0.07$ $e^2/\epsilon\ell$ for $d/\ell=1$ and decreases monotonically as 
interlayer spacing $d$ is increased.)  
Under these circumstances exciton currents are not supercurrents, but 
can still contribute to transport.

Exciton chemical potential gradients, {\it i.e.} differences 
between the electrochemical potential gradients in the top and 
bottom layers, will drive exciton currents.  We choose to 
characterize the counterflow current response by 
an exciton conductivity defined by: 
\begin{equation} 
j_{t}-j_{b} = \sigma^{ex} \; \frac{(E_{t}-E_{b})}{2}
\end{equation}
were $E$ and $j$ are the electrochemical potential gradients and currents, and the 
subscripts $t$ and $b$ refer to the {\em top} and {\em bottom} layers.  
Similarly, charged quasiparticle currents are carried in parallel by the two layers,
sensitive only to the average of the electrochemical potential gradients
in the two layers, and characterized by 
a quasiparticle conductivity:
\begin{equation} 
j_{t}+j_{b}= \sigma^{QP}\; \frac{(E_{t}+E_{b})}{2} 
\end{equation}
It follows that when current flows only in the top layer
\begin{equation} 
E_{t} = (\rho^{QP} + \rho^{ex})j \equiv \rho^{Drive} j 
\end{equation} 
where $\rho^{ex}= (\sigma^{ex})^{-1}$ and $\rho^{QP} = (\sigma^{QP})^{-1}$ 
are the exciton and quasiparticle resistivities.  The drag voltage measured in the 
bottom layers is then 
\begin{equation} 
E_{b} = (\rho^{QP}-\rho^{ex}) j \equiv \rho^{Drag} j.
\end{equation}
When the exciton fluid condenses into a two-dimensional counterflow superfluid,
$\rho^{ex}$ vanishes for small currents and the longitudinal and drag resistivities are identical.
This limit is often closely approached experimentally in bilayer exciton condensates.  
For this reason large positive drag voltages are routinely used as a fingerprint 
of exciton condensates.  
Since the charge gap does not change appreciably, we do not expect that $\rho^{QP}$ will
vary strongly as the $\epsilon_{b1} = \epsilon_{b0}$ boundary of the 
green region of Fig.~\ref{fig:phase_diagram} is approached.
On the other hand, $\rho^{ex}$ becomes finite when the ambient temperature
exceeds $T_{KT}$, and we believe that it can become large as we explain below.

The ground state of density-balanced quantum Hall excitonic superfluids
in the quantum Hall regime can be viewed as a fluid of excitons that 
interact weakly\cite{Lozovik1981} in the limit of small layer separations ($d/\ell < 1$) 
appropriate to graphene-based double layer systems.  In this picture 
both electrons and holes have density $n^{ex} = (\pi \ell^2)^{-1}$.
For a two-dimensional system of interacting bosons the 
superfluid phase stiffness 
\begin{equation} 
\rho = \frac{ \hbar^2 n_s}{2 m^*},
\end{equation} 
where $n_s$ is the boson density and $m^*$ is the particle mass.
Since the exciton density is constant as a function of band 
parameters, it follows that the phase stiffness in double bilayer systems 
vanishes as $\epsilon_{b1} \to \epsilon_{b0}$ from above not because the 
exciton density $n^{ex}$ vanishes, but because the 
exciton mass ($m^* \to m^{ex}$) diverges.
In a simple Drude picture the exciton conductivity
\begin{equation} 
\sigma^{ex} = \frac{n^{ex} e^2 \tau^{ex}}{m^{ex}} = \frac{e^2}{h}  \frac{ 4 \pi \rho \, \tau^{ex}}{\hbar} =  \sigma^{QP}_0 \frac{ \tau^{ex} m^{QP}}{2 \tau^{QP} m^{ex}},
\end{equation}
where $\sigma^{QP}_0$, $m^{QP}$, and $\tau^{QP}$ are the values appropriate 
for the quasiparticle system in the absence of a magnetic field.  
We conclude that at any fixed temperature $\sigma^{ex}$ should become smaller than 
$\sigma^{QP}$ when $m^{ex}$ diverges as $\epsilon_{b1} \to \epsilon_{b0})$,
and the drag resistance should become large and negative.  

\begin{figure}[t]
	\centering
	\includegraphics[width=0.99\columnwidth]{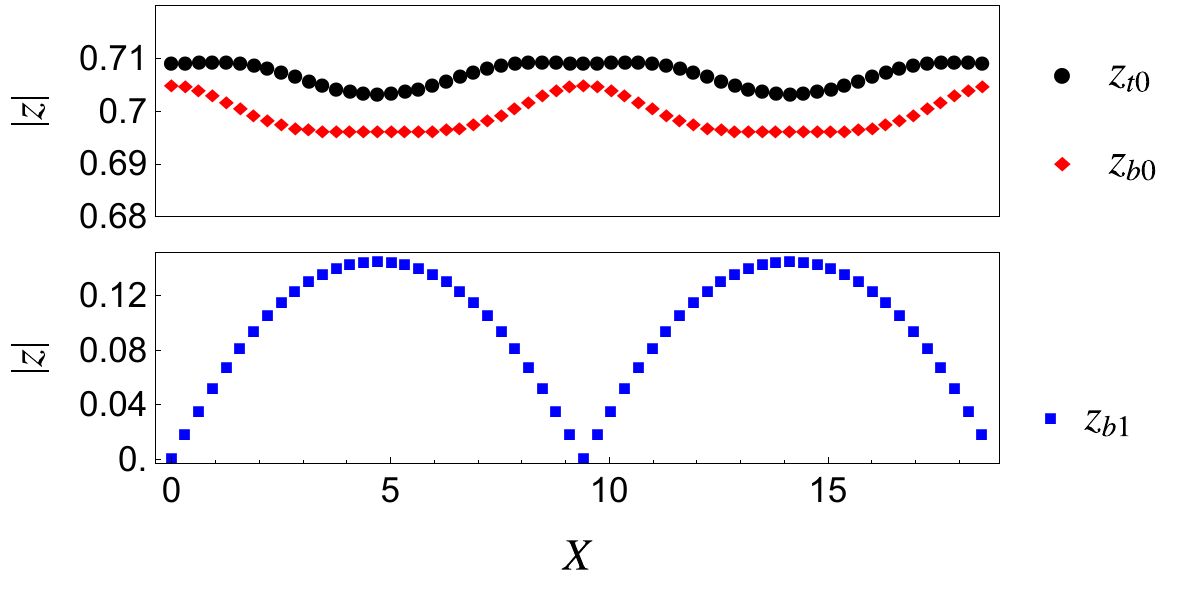}
	\caption{\label{fig:stripe} Layer and orbital projected wavefunction of the stripe phase at $\epsilon_{t0}=\epsilon_{b0}=0$ and $\epsilon_{b1}=-0.016$. }
\end{figure}

\emph{Stripe phase instability.}---\nobreakdash
When $\epsilon_{b1}<\epsilon_{b0}$
the phase energy kernel has the form $ - A q^2 + B q^4$, 
becoming negative over a finite range of small $q$ values
as shown in the inset of Fig.~\ref{fig:kernel}.
To identify the nature of the states implied by this instability, 
we have performed self-consistent mean field calculations that allow 
translational symmetry to be broken along one direction, taken
be the $x$ direction.  We find that the resulting stripe states 
have their lowest energies when their periods in $x$  are 
close to $2 \pi/q^*$ where $q^*$ is the value of $q$ at which the 
phase kernel eigenvalue reaches its minimum,
Fig.~\ref{fig:stripe} illustrates the variation in the guiding center 
spinors, whose orbital content corresponds to the eigenvector 
of the negative eigenvalue phase mode.  As one passes through the
stripe state red region of Fig.~\ref{fig:phase_diagram} from right to left, the $n=1$ orbital 
content of the wave function increases. 

\emph{Discussion.}---\nobreakdash
In systems of fermions with attractive effective interactions, it is known that the 
crossover from the BCS-theory weak interaction limit, to the BEC limit of weakly 
interacting boson limit is smooth.  The microscopic physics of condensed electron-hole pairs in the 
quantum Hall regime is distinct from the familiar BCS-BEC crossover paradigm because of the 
way in which Landau quantization cuts off the many-particle Hilbert space.   The elementary 
excitation spectrum at long wavelengths consists only of bosonic collective modes formed 
by electron-hole pairs that are more and more weakly bound as wavelengths shorten.
Although strong-positive drag signals suggesting excitionic superfluidity have 
been observed regularly, there have been few observations that signal 
transport contributions from uncondensed bosonic excitations.  
In this Letter we have shown that the superfluid phase stiffness of 
double bilayer quantum Hall exciton condensates vanishes as 
$n=0$ and $n=1$ orbitals states approach degeneracy in one of the layers.
We associate the increase in stiffness with a diverging exciton mass which, we 
argue, will also lead to a diverging excitonic counterflow resistivity and 
to large negative drag resistivities.  Indeed large negative drag 
signals do\cite{Experiments} 
appear in double bilayer graphene near narrow regions of gate voltage settings close to 
orbital degeneracy conditions.  If our interpretation of the drag anomalies is 
correct, the sign of the drag is determined by a competition between 
quasiparticle resistivities that diverge as $T \to 0$ for any gate setting,
and exciton resistivities that diverge at any temperature as gate settings are tuned to 
orbital degeneracy.

\emph{Acknowledgment.}---\nobreakdash
AHM and MX were supported by DOE BES under Award DE-FG02-02ER45958
and by the Welch Foundation under grant TBF1473.

\newpage
.
\newpage
\appendix
% Supplemental material will be separated from main text at submission.
\section{Supplemental material}
\label{supple}

\section{I. Total energy functional and fluctuation kernel matrix}
We derive the total energy functional of fluctuations around mean field ground states
and the fluctuation kernel matrix elements.
For given band parameters $\epsilon_{\alpha,n}$, self-consistent mean field theory 
predicts eigenstate wavefunctions $\{a^m_{\alpha, n}\}$ for both the
ground state $(m=0)$ and the excited states $(m=1,2)$.
The many-body wavefunction of collective fluctuations are given by Eq. 5 where the excited states are mixed into the ground state wavefunction
leading to the instantaneous Slater determinant states.
The total energy functional can be written as
\begin{align}
	\mathcal{E}[\{z, z^*\}] = &\mathcal{E}_0[\{z, z^*\}]
	+\mathcal{E}_{H}[\{z, z^*\}]
	+\mathcal{E}_{F}[\{z, z^*\}]
\end{align}
where
\begin{align}
	\mathcal{E}_0[\{z, z^*\}] = \sum_{m', m, X} \sum_{\alpha,n}\epsilon_{\alpha,n}a^{*m'}_{\alpha,n}a^{m}_{\alpha,n} z^*_{m',X}z_{m,X}
\end{align}
is the single-particle energy and
\begin{widetext}
	\begin{align}
	% Hartree total energy
	\mathcal{E}_H[\{z, z^*\}] &= 
	\frac{1}{2S} \sum_{\substack{X_1,X_2\\q_x}}\sum_{\alpha,\beta} V_{\alpha\beta}(q_x) e^{iq_x(X_1-X_2)}
	\sum_{m_1,m_1'} \mathcal{F}^{\alpha}_{m_1m_1'}(q_x)z^*_{m_1, X_1}z_{m_1',X_1}
	\sum_{m_2,m_2'} \mathcal{F}^{\beta}_{m_2m_2'}(-q_x)z^*_{m_2, X_2}z_{m_2',X_2} \\
	% Fock total energy
	\mathcal{E}_F[\{z, z^*\}] &= 
	-\frac{1}{2S}\sum_{\substack{X_1,X_2\\\bm{q}}}\sum_{\alpha,\beta} V_{\alpha\beta}(\bm{q}) \delta_{q_y\ell^2, X_1-X_2}
	\sum_{m_1,m_2'} \mathcal{F}^{\alpha}_{m_1m_2'}(\bm{q})z^*_{m_1, X_1}z_{m_2',X_2}
	\sum_{m_2,m_1'} \mathcal{F}^{\beta}_{m_2m_1'}(-\bm{q})z^*_{m_2, X_2}z_{m_1',X_1}
	\end{align}
\end{widetext}
are interaction total energies, where the subscripts $H$ and $F$ stand for Hartree and Fock components, respectively.
We have introduced the modified form factors
\begin{align}
	\mathcal{F}^{\alpha}_{mm'}(\bm{q}) = \sum_{n,n'}a^{*m'}_{\alpha,n}a^{m}_{\alpha,n} F_{nn'}(\bm{q})
\end{align}
where $F_{nn'}(\bm{q})$ is the familiar two-dimensional gas Landau-level form factor given by
%\begin{widetext}
	\begin{align}
	F_{nn'}(\bm{q}) =& \left(\frac{n_<!}{n_>}\right)^{\frac{1}{2}} \left\{\frac{[\text{sgn}(n-n')q_y+iq_x]\ell}{\sqrt{2}}\right\}^{n_>-n_<} \notag\\
	&\times 
	L_{n_<}^{n_>-n_<}\left(\frac{q^2\ell^2}{2}\right) \exp\left(-\frac{q^2\ell^2}{4}\right)
	\end{align}
%\end{widetext}
where $n_>(n_<)$ is the greater (lesser) one of $n$ and $n'$. $L_n^{n'}(x)$ is the generalized Laguerre polynomial and $\text{sgn}(x)$ the sign function.

The fluctuation kernel is the coefficient of second order expansion of the total energy around the mean field ground state.
For small amplitude fluctuations, i.e. $|z_{m=1,2}|\ll 1$ and $|z_{m=0}|\approx 1$,
the normalization condition up to second order in the amplitudes can be expressed as
\begin{align}
	z_{0,X} \approx 1 - \frac{1}{2}|z_{1,X}|^2-\frac{1}{2}|z_{2,X}|^2
	\label{sqrtexpansion}
\end{align}
where we have chosen the phase of $z_{0,X}$ to be zero anywhere.
By substituting Eq. \ref{sqrtexpansion}, we can rewrite the total energy functional 
which then depends only on $Z_{i,X}=\{z_{1,X}, z^*_{1,X}, z_{2,X}, z^*_{2,X}\}$ where $i=1,...4$.
By taking second order derivative with respect to $Z_{i,X}$ and Fourier transformation in guiding center coordinate $X$,
we obtain the following independent fluctuation kernel matrix elements.
\begin{widetext}
For $i=2,4$ and $j=1,3$,
\begin{align}
%%		\frac{\partial^2 \mathcal{E}[\{z,z^*\}]}{\partial z^*_{m}\partial z_{m'}} (q)
\mathcal{K}_{ij}(q)
		= &\sum_{\alpha, n} \epsilon_{\alpha n}(a^{*m}_{\alpha n}a^{m'}_{\alpha n}-a^{*0}_{\alpha n}a^{0}_{\alpha n}\delta_{mm'}) \notag\\
		   &+\frac{1}{L_y} \sum_{\substack{n_1n_1'\\n_2n_2'}} \sum_{\alpha\beta}
		         H^{\alpha\beta}_{n_1n_1'n_2n_2'}(0) (a^{*m}_{\alpha n_1}a^{m'}_{\alpha n_1'}-a^{*0}_{\alpha n_1}a^{0}_{\alpha n_1'}\delta_{mm'})
		                   a^{*0}_{\beta n_2}a^{0}_{\beta n_2'}
                +H^{\alpha\beta}_{n_1n_1'n_2n_2'}(q)a^{*m}_{\alpha n_1}a^{0}_{\alpha n_1'}a^{*0}_{\beta n_2}a^{m'}_{\beta n_2'} \notag\\
          &-\frac{1}{L_y} \sum_{\substack{n_1n_1'\\n_2n_2'}} \sum_{\alpha\beta}
               I^{\alpha\beta}_{n_1n_1'n_2n_2'}(0)  (a^{*m}_{\alpha n_1}a^{m'}_{\beta n_2'}-a^{*0}_{\alpha n_1}a^{0}_{\beta n_2'}\delta_{mm'})
                           a^{*0}_{\alpha n_1'} a^{0}_{\beta n_2}         
              -I^{\alpha\beta}_{n_1n_1'n_2n_2'}(q)a^{*m}_{\alpha n_1}a^{m'}_{\alpha n_1'}a^{*0}_{\beta n_2}a^{0}_{\beta n_2'},  \label{k21}\\
\intertext{and for $i,j=1,3$, we have}
%          \frac{\partial^2 \mathcal{E}[\{z,z^*\}]}{\partial z_{m}\partial z_{m'}} (q)
\mathcal{K}_{ij}(q)
          =&\frac{1}{L_y} \sum_{\substack{n_1n_1'\\n_2n_2'}} \sum_{\alpha\beta}
                    H^{\alpha\beta}_{n_1n_1'n_2n_2'}(q)a^{*0}_{\alpha n_1}a^{m}_{\alpha n_1'}a^{*0}_{\beta n_2}a^{m'}_{\beta n_2'} 
            -\frac{1}{L_y} \sum_{\substack{n_1n_1'\\n_2n_2'}} \sum_{\alpha\beta}
                     I^{\alpha\beta}_{n_1n_1'n_2n_2'}(q)a^{*m}_{\alpha n_1}a^{0}_{\alpha n_1'}a^{*m'}_{\beta n_2}a^{0}_{\beta n_2'}.     \label{k24}                         
\end{align}
\end{widetext}
where $H$ and $I$ are interaction integrals defined as
%\begin{widetext}
\begin{align}
    &H^{\alpha\beta}_{n_1n_1'n_2n_2'}(q) = \frac{1}{2\pi\ell^2} V_{\alpha\beta}(q)F_{n_1n_1'}(q)F_{n_2n_2'}(-q) \\
	&I^{\alpha\beta}_{n_1n_1'n_2n_2'}(q) = \int \frac{d\bm{p}}{(2\pi)^2} e^{iqp_y\ell^2} V_{\alpha\beta}(\bm{p})F_{n_1n_1'}(\bm{p})F_{n_2n_2'}(-\bm{p})
\end{align} 
%\end{widetext}
The first line of Eq.~\ref{k21} is the single-particle term whereas the second and third lines corresponds to Hartree and Fock interaction contributions, respectively.
In contrast, Eq.~\ref{k24} has only interaction terms, the first and the second of which correspond to Hartree and Fock terms, respectively.
The rest of the matrix elements can be found by Hermitian conjugation relation $\mathcal{K}_{ij}(q)=\mathcal{K}^*_{ji}(q)$.

\section{II. Eigenenergies of the Fluctuation Kernel $\mathcal{K}(q)$}
In this section, we present examples of the quadratic fluctuation and 
collective mode energies for several different regions of the double-bilayer 
phase diagram illustrated in Fig.~\ref{fig:phase_diagram}.

\subsection{II. A. Fully Polarized in Top Layer}

\begin{figure}[h!]
	\centering
	\includegraphics[width=0.9\columnwidth]{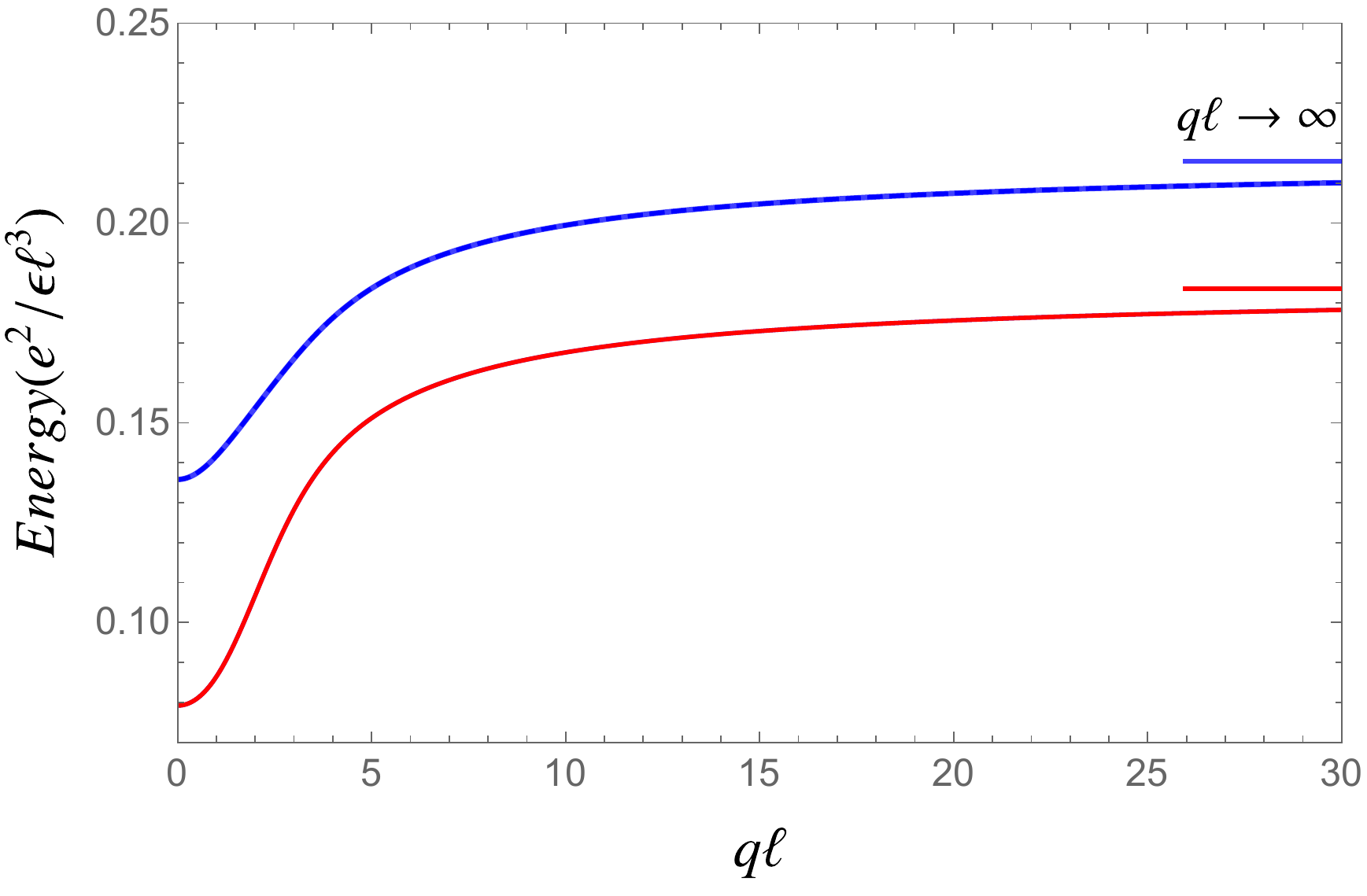}
	\caption{\label{SFIG_1} The fluctuation kernel $\mathcal{K}(q)$ at $\epsilon_{b0}=0.9$ and $\epsilon_{b1}=1.1$. 
	Each curve is two-fold degenerate. At $q=0$, the red lines correspond to
	 $z_{1,q},\ z^*_{1,-q}$ fluctuations and the  
	blue lines to $z_{2,q},\ z^*_{2,-q}$ fluctuations.
	For $q \neq 0$, $z_{1,q}$ mixes with $z_{2,q}$ and $z^*_{1,-q}$ mixes with $z^*_{2,-q}$.
	The solid lines indicate the $q \ell \to \infty$.}
\end{figure}

When the top layer Landau $n=0$ is lowest in energy by a sufficiently large 
margin, the ground state is both layer and orbital polarized to the
$n=0$ orbital of the top layer.  
For example for $\epsilon_{b0}=0.9$ and $\epsilon_{b1}=1.1$,
the mean field eigenstates are (ascending order in eigenenergy): 
\begin{align}
	&|m=0\rangle = |t,n=0\rangle, \notag \\
	&|m=1\rangle = |b,n=0\rangle, \notag \\
	&|m=2\rangle = |b,n=1\rangle.
\end{align}
The fluctuation kernel  which specifies the energies of 
transitions from the $|m=0\rangle$ level to the
 $|m=1\rangle$ and $|m=2\rangle$ levels is plotted in Fig.\ref{SFIG_1}.
 
\subsection{II. B. Ising Quantum Hall Ferromagnet}
In this region mean-field predicts ground state that are mixtures of
$|t, n=0\rangle$ and $|b, n=1\rangle$.
We take, for example, the single particle energies 
$\epsilon_{b0}=1$ and $\epsilon_{b1}=0.1$.
The mean field eigenstates are
\begin{align}
&|m=0\rangle = 0.90 |t,n=0\rangle + 0.43|b,n=1\rangle, \notag \\
&|m=1\rangle = 0.43 |t,n=0\rangle - 0.90|b,n=1\rangle, \notag \\
&|m=2\rangle = |b,n=0\rangle.
\end{align}
\begin{figure}[h!]
	\centering
	\includegraphics[width=0.9\columnwidth]{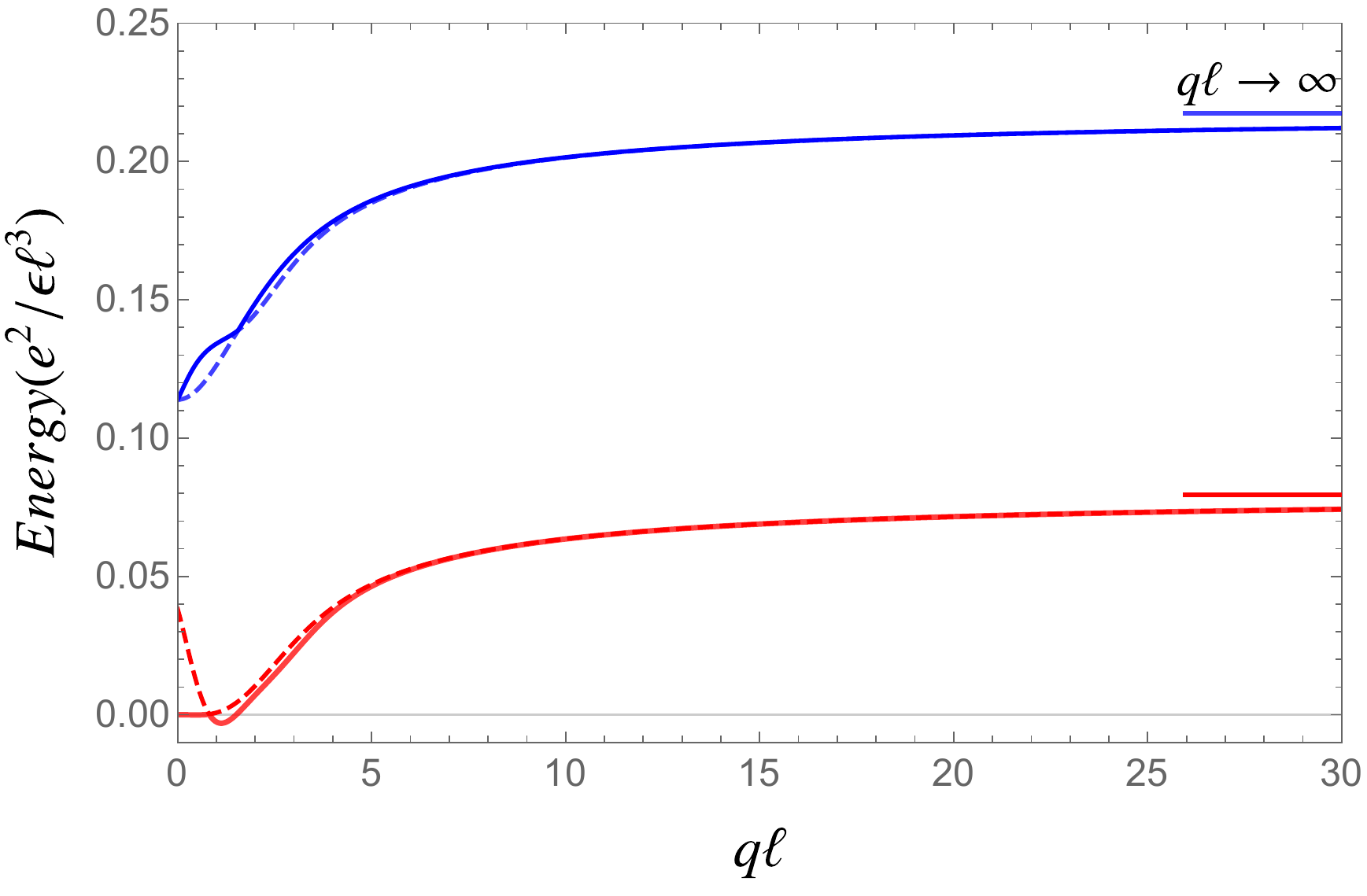}
	\caption{\label{SFIG_2} Eigenenergy of fluctuation kernel $\mathcal{K}(q)$ at $\epsilon_{b0}=1$ and $\epsilon_{b1}=0.1$. 
		At $q=0$, red lines correspond to linear superposition modes of $z_{1,q},\ z^*_{1,-q}$ and 
		blue lines to linear superposition modes of $z_{2,q},\ z^*_{2,-q}$. For $q \neq 0$,  each eigenstate is linear combination of all modes.}
\end{figure}
The eigenenergy of fluctuations to $|m=1\rangle$ and $|m=2\rangle$ states is plotted in Fig.\ref{SFIG_2}.
The negative eigenenergy shows that the mean-field ground state in this region is not stable.

\subsection{II. C. $n=1$ Layer and Orbitally Polarized state}
\begin{figure}[h!]
	\centering
	\includegraphics[width=0.9\columnwidth]{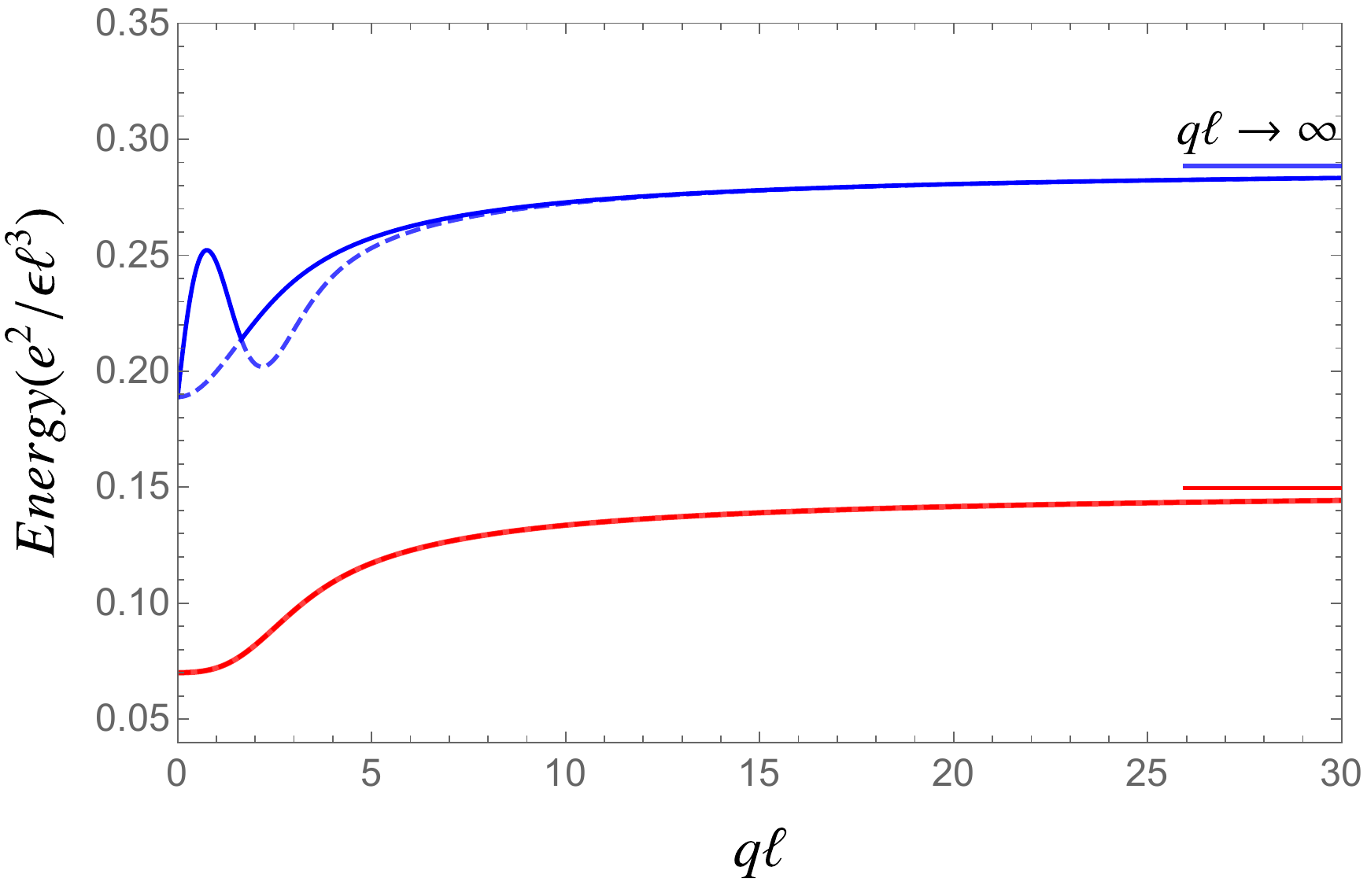}
	\caption{\label{SFIG_3} Eigenenergy of fluctuation kernel $\mathcal{K}(q)$ at $\epsilon_{b0}=0.5$ and $\epsilon_{b1}=-1$. 
		Red line is two fold degenerate and correspond to degenerate modes of $z_{1,q},\ z^*_{1,-q}$ and 
		blue lines to modes $(z_{2,q}\pm\ z^*_{2,-q})/\sqrt{2}$. }
\end{figure}

 In the region where ground state is both layer and orbital polarized to $|b, n=1\rangle$.
We set, for example,
$\epsilon_{b0}=0.5$ and $\epsilon_{b1}=-1$.
The mean field eigenstates are
\begin{align}
&|m=0\rangle = |b,n=1\rangle, \notag \\
&|m=1\rangle = |t,n=0\rangle, \notag \\
&|m=2\rangle = |b,n=0\rangle.
\end{align}
The eigenenergy of fluctuations to $|m=1\rangle$ and $|m=2\rangle$ states is plotted in Fig.\ref{SFIG_3}.

\subsection{II. D. Layer Polarized Mixed Orbital State} 
We consider the region where mean-field ground state is mixture of $|b,n=0\rangle$ and $|b,n=1\rangle$.
As an example, we set $\epsilon_{b0}=-1$ and $\epsilon_{b1}=-1.3$.
The mean field eigenstates are
\begin{align}
&|m=0\rangle = 0.21 |b,n=0\rangle + 0.98|b,n=1\rangle, \notag \\
&|m=1\rangle = 0.98 |b,n=0\rangle - 0.21|b,n=1\rangle, \notag \\
&|m=2\rangle = |t,n=0\rangle.
\end{align}

\begin{figure}[ht!]
	\centering
	\includegraphics[width=0.9\columnwidth]{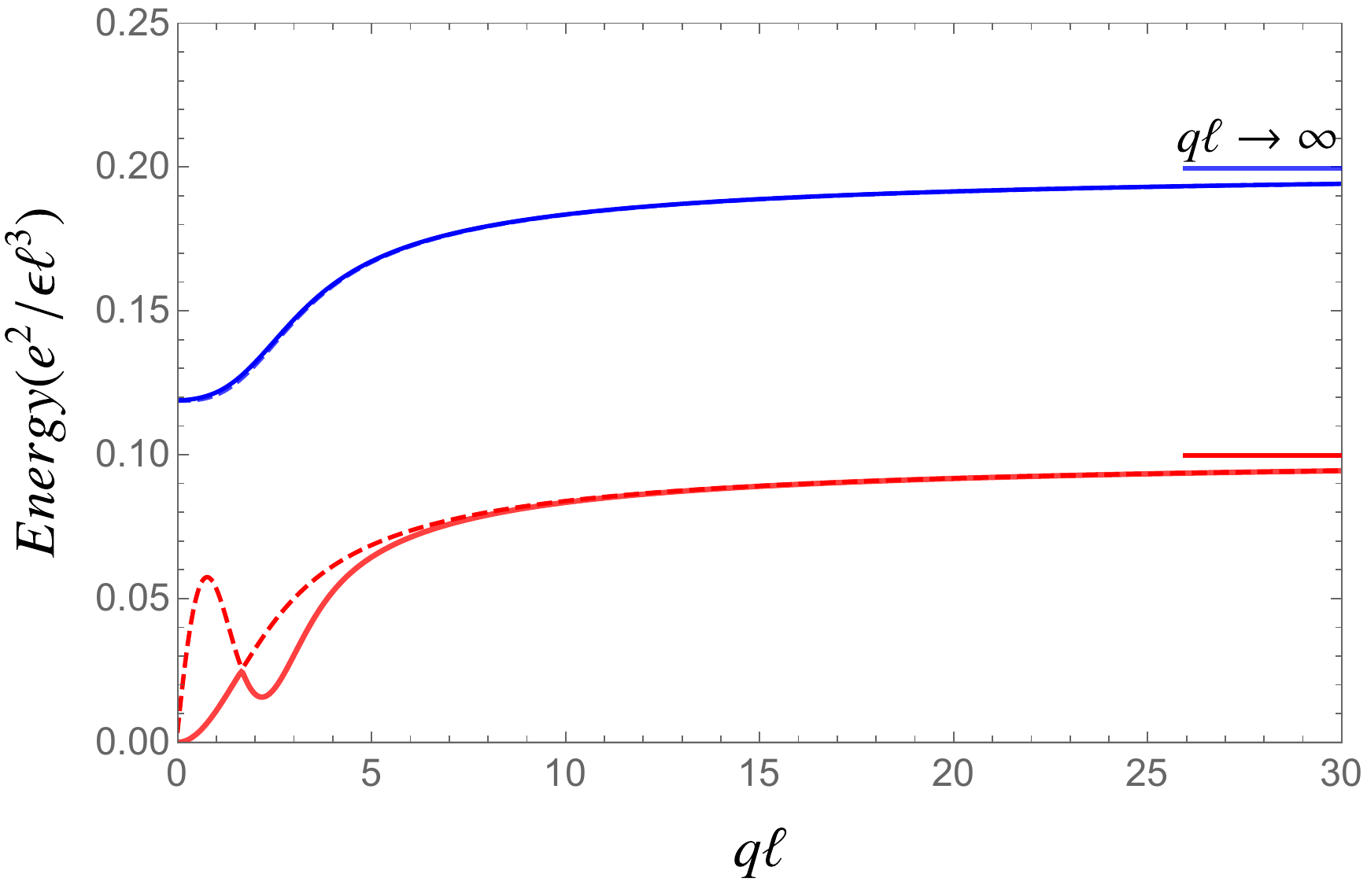}
	\caption{\label{SFIG_4} Eigenenergy of fluctuation kernel $\mathcal{K}(q)$ at $\epsilon_{b0}=-1$ and $\epsilon_{b1}=-1.3$. 
		Red lines correspond to linear superposition of modes $z_{1,q}, \ z^*_{1,-q}$ and
		blue lines are two fold degenerate and correspond to degenerate modes of $z_{2,q},\ z^*_{2,-q}$.}
\end{figure}
The eigenenergy of fluctuations to $|m=1\rangle$ and $|m=2\rangle$ states is plotted in Fig.\ref{SFIG_4}.

\subsection{II. E. Layer and orbital polarized state}
\begin{figure}[h!]
	\centering
	\includegraphics[width=0.9\columnwidth]{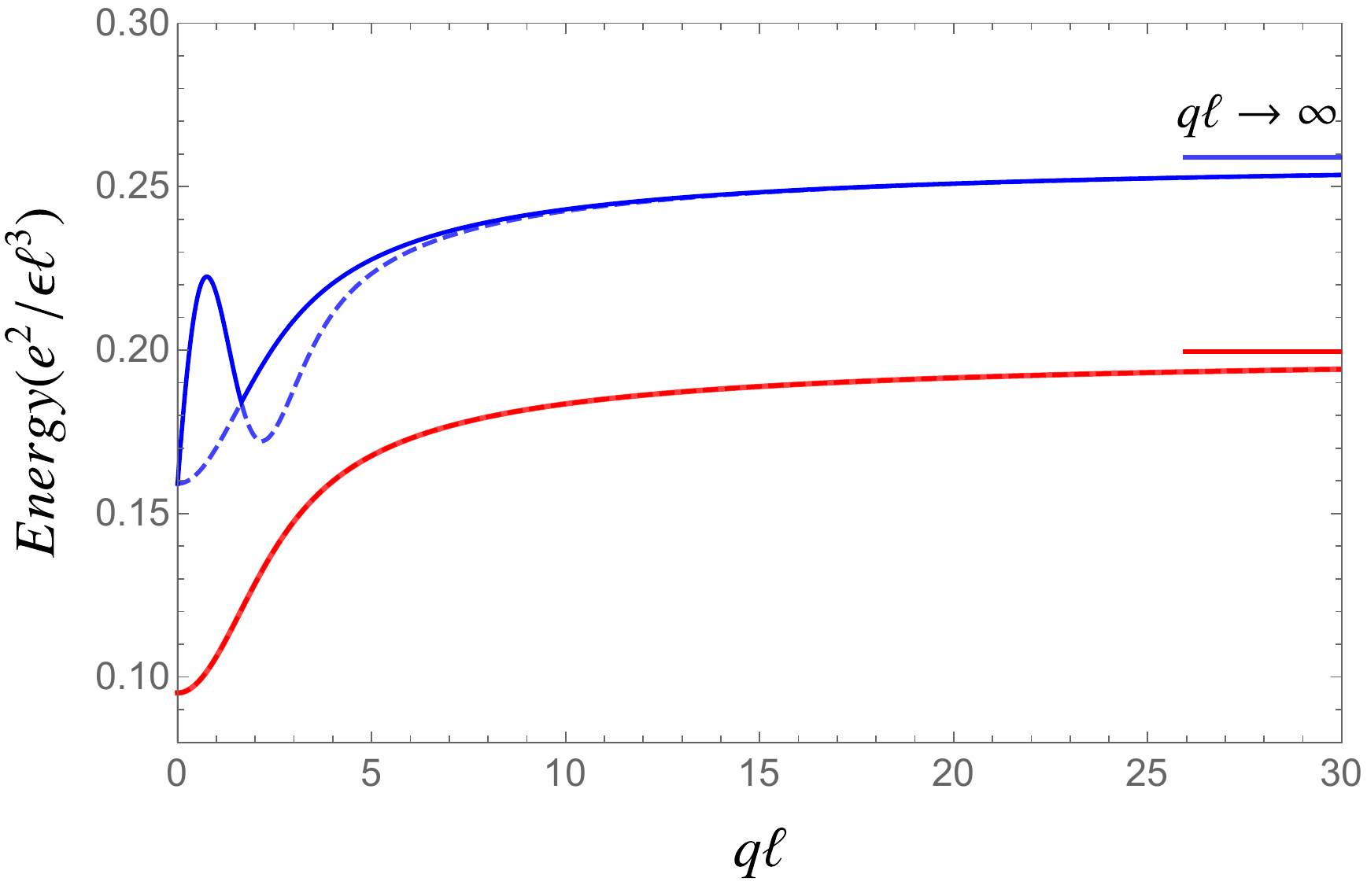}
	\caption{\label{SFIG_5} Eigenenergy of fluctuation kernel $\mathcal{K}(q)$ at $\epsilon_{b0}=-1$ and $\epsilon_{b1}=0$. 
		Red line is two fold degenerate and correspond to degenerate modes of $z_{1,q},\ z^*_{1,-q}$ and 
		blue lines to modes $(z_{2,q}\pm\ z^*_{2,-q})/\sqrt{2}$. }
\end{figure}

Setting $\epsilon_{b0}=-1$ and $\epsilon_{b1}=0$ leads to the mean field eigenstates:
\begin{align}
&|m=0\rangle = |b,n=0\rangle, \notag \\
&|m=1\rangle = |t,n=0\rangle, \notag \\
&|m=2\rangle = |b,n=1\rangle.
\end{align}
The eigenenergy of fluctuations to $|m=1\rangle$ and $|m=2\rangle$ states is plotted in Fig.\ref{SFIG_5}.

\subsection{II. F. ``Mixing all" region}
In this region, mean field calculation predicts ground state to be a mixture of 
$|t,n=0\rangle$, $|b,n=0\rangle$ and $|b,n=1\rangle$.
As an example, we set $\epsilon_{b0}=0$ and $\epsilon_{b1}=-0.2$.
\begin{figure}[t!]
	\centering
	\includegraphics[width=0.9\columnwidth]{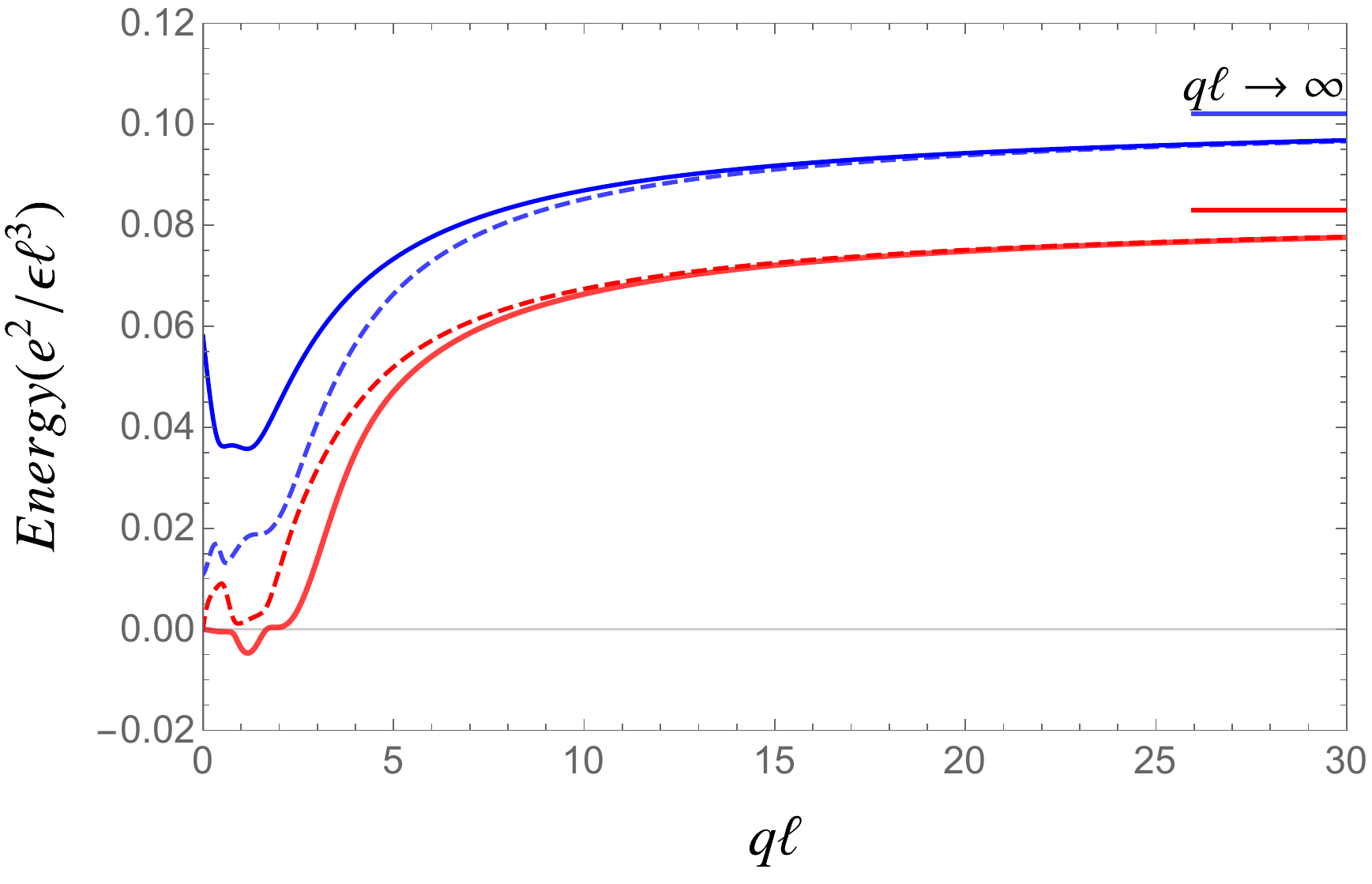}
	\caption{\label{SFIG_6} Eigenenergy of fluctuation kernel $\mathcal{K}(q)$ at $\epsilon_{b0}=0$ and $\epsilon_{b1}=-0.2$. 
		Each eigenstate branch is a mixture of all four modes $z_{1,q},\ z^*_{1,-q}, z_{2,q},\ z^*_{2,-q}$.}
\end{figure}

The eigenenergy of fluctuations to $|m=1\rangle$ and $|m=2\rangle$ states is plotted in Fig.\ref{SFIG_6}.
As expected, this mean field ground state is not stable.
\newline

\end{document}